\documentclass{mem}
\usepackage{natbib}\usepackage{txfonts}\usepackage{balance}
\usepackage{graphicx}
\usepackage[a4paper]{hyperref}
\idline{00}{0}
\def \th {\thinspace}

\def\approxgt{\mathrel{\hbox{\rlap{\lower.55ex \hbox {$\sim$}} \kern-.3em \raise.4ex \hbox{$>$}}}}
\def\lesssim{\mathrel{\hbox{\rlap{\lower.55ex \hbox {$\sim$}} \kern-.3em \raise.4ex \hbox{$<$}}}}
\def\approxlt{\mathrel{\hbox{\rlap{\lower.55ex \hbox {$\sim$}} \kern-.3em \raise.4ex \hbox{$<$}}}}

\begin{document}
\title{The nature of the 4th track in GX\th 5-1: discovery of Fe XXVI RRC in massive flares}
\subtitle{}

\author{
M. J.\,Church\inst{1,2} 
\and O.\, Dimbylow\inst{1}
\and C.\, Peach\inst{1}
\and M.\, Balucinska-Church\inst{1,2}
}

\offprints{M. J. Church}

\institute{School of Physics \& Astronomy,
University of Birmingham, Birmingham B15 2TT, UK
\and
Astronomical Observatory, Jagiellonian University,
ul. Orla 171, 30-244, Cracow, Poland
\email{mjc@star.sr.bham.ac.uk}}

\authorrunning{Church et al.}
\titlerunning{The 4th track}

\abstract{
We present an explanation of the 4th branch of the Z-track based on analysis of high-quality 
{\it RXTE} data on the source GX\th 5-1. Spectral analysis shows that the physical evolution 
on the 4th track is a continuation of the flaring branch which we previously proposed consists 
of unstable nuclear burning of the accretion flow on the neutron star. In flaring there is a 
huge increase of the neutron star emission from a volume that increases to a radius of 21 km. 
The 4th branch is shown to consist of flaring under conditions that the mass accretion rate and 
thus the total source luminosity is falling. We detect strong emission on the flaring and 4th 
branches at energies between 7.8 - $\sim$9.4 keV inconsistent with origin as Fe K emission, which we 
suggest is the radiative recombination continua (RRC) of iron Fe XXVI at 
9.28 keV and of lower states. Evolution of the emission takes place, the energy falling but the
flux increasing strongly, consistent with production in the large volume of unstable nuclear 
burning around the neutron star which eventually cools.
}
\maketitle{}

\section{Introduction}

The Z-track sources form the brightest group of Low Mass X-ray Binary (LMXB) sources containing 
a neutron star, including: GX\th 340+0, GX\th 5-1, Cyg\th X-2, Sco\th X-1, GX\th 17+2 
and GX\th 349+2 (Hasinger \& van der Klis 1989). All of these persistently radiate at or above
the Eddington limit for a neutron star. The sources trace a Z-shaped pattern in X-ray
hardness versus intensity consisting of the horizontal branch (HB), the normal branch (NB) and 
the flaring branch (FB). The sources are divided into the Cyg-like sources: GX\th 340+0, GX\th 5-1
and \hbox{Cyg\th X-2} with long horizontal branches and the Sco-like sources: Sco\th X-1, GX\th 17+2
and GX\th 349+2 with short horizontal branches but stronger flaring branches. This strong 
hardness evolution along the Z-track is clearly 
due to major physical changes taking place within the sources, the nature of which has not 
been understood. 
\begin{figure*}                                                           
\includegraphics[width=44mm,height=135mm,angle=270]{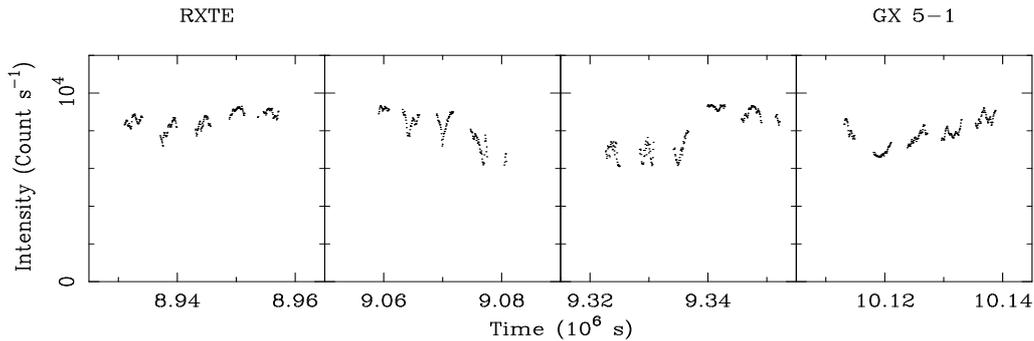} 
\caption{Lightcurve of the observation in the energy band 1.9 - 18.5 keV. The data consist of 4 sub-observations
so that, to improve clarity, the gaps between the sub-observations have been compressed.}
\end{figure*}
However, based on analysis of {\it Rossi-XTE} data on the sources GX\th 340+0, 
GX\th 5-1 and Cygnus\th X-2, we have proposed a model of the Cyg-like sources (Church et al. 2006; 
Jackson et al. 2008, Ba\l uci\'nska-Church et al. 2009). In this model, the increase in X-ray 
intensity moving on the NB from the soft apex to the hard apex is due to an increase 
of mass accretion rate. There is a resultant heating of the neutron star and
a contraction in the emitting area so that the flux emitted per unit area of the 
neutron star surface becomes super-Eddington leading to disruption of the inner disk
and the launching of jets detected as radio emission on this part of the Z-track.
On the FB, spectral fitting showed little change of $\dot M$ but the large increase in the
luminosity of the neutron star indicated unstable nuclear burning on the neutron star.
Moreover there was good agreement at the onset of flaring (the soft apex) between the mass 
accretion rate per unit area of the neutron star $\dot m$ and the theoretical value below
which burning of the accretion flow on the neutron star becomes unstable.

In addition, the Cyg-like sources sometimes display extra unusual 
tracks which similarly have not been understood. In Fig. 2 we show the additional 
4th branch in GX\th 5-1 in the observation analysed in the present work.
It can be seen that this begins at the end of the FB and moves in the direction of decreasing 
hardness and decreasing intensity. It has been referred to as the ``dipping track'' because of
the decreasing intensity (e.g. Kuulkers \& van der Klis 1996) by analogy with the dipping class of LMXB
in which absorption in the bulge in the outer disk causes orbital-related
decreases of intensity. This absorption may cause a hardening of the spectrum, but energy-independent
dipping (Church \& Ba\l uci\'nska-Church 1993) and also softening (Church \& Ba\l uci\'nska-Church 
1995) in dips is also observed. This range of behaviour may be explained in terms of two continuum 
emission components within the sources (Church \& Ba\l uci\'nska-Church 1995). The apparent
dipping in a Z-track source led to the suggestion that the Cyg-like sources have high 
inclination and that this explains the difference between the Cyg-like and Sco-like sub-groups. 
It should be noted, however, that we would not expect the 4th track 
to be associated with flaring but to take place on any part of the Z-track.

In the present work, we carry out an investigation of the 4th track in GX\th 5-1.

\section{Observations and analysis}

The observations analysed were made using {\it RXTE} in October and November 1996.
The lightcurve of these observations is shown in Fig. 1 and the corresponding
Z-track is shown in Fig. 2 in which the hardness ratio is
obtained by the division of light curves in the bands 10.2 - 17.5 and 6.5 - 10.2 keV.
Spectra were accumulated as a function of position along the
Z-track by defining boxes in hardness-intensity typically 0.005 wide in hardness and 100 count 
s$^{-1}$ wide in intensity. Two spectra were chosen on the HB, one at the hard apex, two on the NB, one 
at the soft apex, two on the FB, one at the 3rd apex and three others on the 4th track. 

\begin{figure}[!ht]                          
\includegraphics[width=54mm,height=64mm,angle=270]{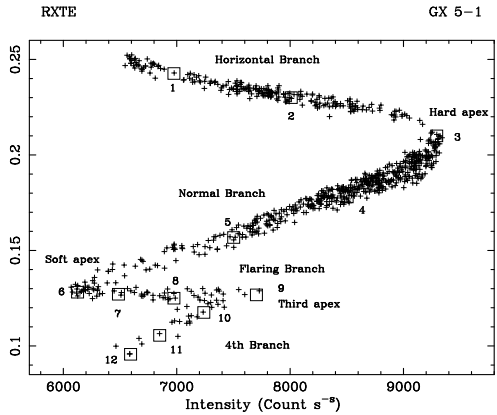}
\caption{Z-track of the observations analysed showing the fourth track in GX\th 5-1.}
\end{figure}

In our previous work on the three Cygnus\th X-2 like sources (Church et al. (2006); Jackson et al. 
(2008); Ba\l uci\'nska-Church et al. 2009) we showed that spectral evolution along the Z-track
was well-described by application of the Extended ADC emission model for low mass X-ray binaries
(Church \& Ba\l uci\'nska-Church 2004) in which the continuum emission comprises the blackbody 
emission of the neutron star and the dominant Comptonized emission of the extended accretion
disk corona (ADC). Application of this model suggested a clear physical explanation of the 
Z-track as summarised in Sect. 1. Thus this model was applied 
in the present work in the form {\sc ab}$\ast$ {\sc (bb + cut)} where {\sc bb} is the blackbody 
emission of the neutron star, {\sc cut} is a cut-off power law description of Comptonization is an 
extended accretion disk corona (Church \& Ba\l uci\'nska-Church 2004) and {\sc ab} is absorption. 
This model provided good fits on the HB and NB, however, on the flaring branch and in particular 
on the 4th branch, strong, broad line emission became apparent at energies above 9 keV. This was 
modelled using the {\sc redge} model in {\it Xspec} for the recombination radiation continuum (RRC) 
emitted following recombination of highly ionized states of ions, particularly of iron, since such 
transitions from the continuum to the ground state have energies above 9 keV. 

\section{Results}

In Fig. 3 we show that part of the lightcurve of the source during which the source moves
on the 4th branch.
Horizontal lines are drawn 
\begin{figure}[!h]                                                           
\includegraphics[width=54mm,height=64mm,angle=270]{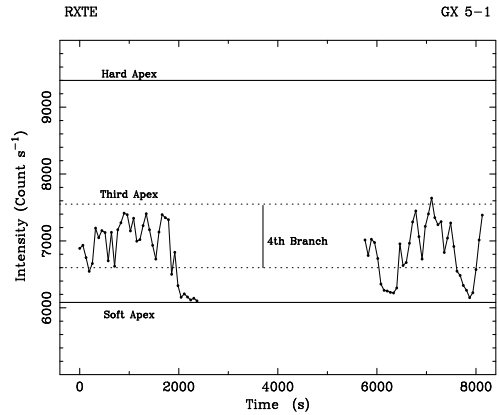}
\caption{The part of the lightcurve during which the source moves on the 4th branch.}
\end{figure}
at the intensities of the soft apex, the third
apex and the end of the 4th branch. In the
second block of data it can be seen that the source moves from the soft apex
in strong flaring up to the third apex. It then moves several times between the third 
apex and the end of the 4th branch eventually returning to the soft apex. We will show that although the
intensity decreases from the 3rd apex to the end of the 4th branch, the neutron star luminosity 
in fact {\it increases strongly} so that the 4th branch actually consists of a continued strong
increase of flare intensity, i.e. of the neutron star luminosity.
\begin{figure}                                                           
\includegraphics[width=54mm,height=64mm,angle=270]{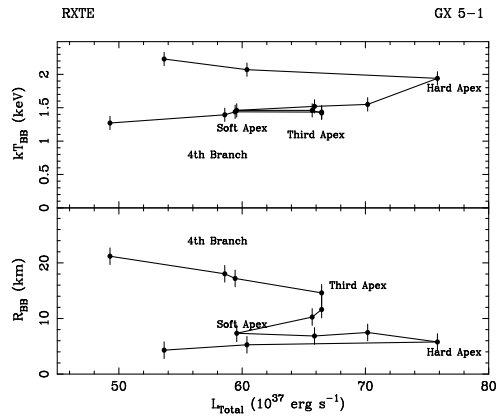}
\caption{Evolution of neutron star blackbody temperature (upper panel) and blackbody radius 
(lower panel) along the Z-track.}
\end{figure}

Spectral fitting results are shown in Fig. 4 for the neutron star blackbody
temperature $kT_{\rm BB}$ (upper panel) and blackbody radius $R_{\rm BB}$ (lower panel)
as a function of the total luminosity in the band 1 - 30 keV. The behaviour on the NB and HB
is the same as found by Jackson et al. (2008), i.e. on the NB, $kT_{\rm BB}$
increases from a low value at the soft apex (1.5 keV)
to a high value on the HB end of the Z-track (2.2 keV). At the same time the blackbody
radius $R_{\rm BB}$ decreases from a value approaching the radius of the neutron star
to a much smaller value. On the FB and the 4th branch, the behaviour of
both $kT_{\rm BB}$ and $R_{\rm BB}$ is monotonic, the major effect being the increase of $R_{\rm BB}$.

The difference between the FB and the 4th branch is clearly the strong decrease in total luminosity
on the 4th branch. The results are more clearly seen if we show them separately from this
decrease of $L_{\rm Tot}$ which is done by re-plotting as a function of the blackbody luminosity
$L_{\rm BB}$ as shown in Fig. 5.

\begin{figure}                                                           
\includegraphics[width=54mm,height=64mm,angle=270]{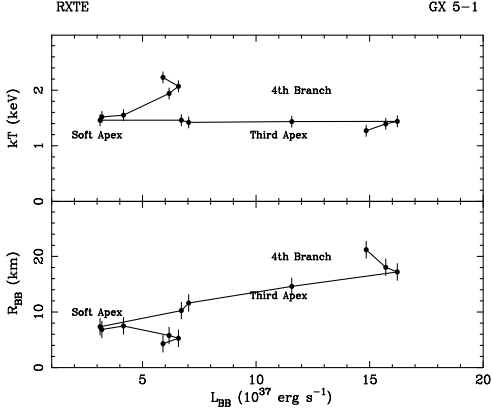}
\caption{The unfolded evolution of $kT_{\rm BB}$ (upper panel) and $R_{\rm BB}$ (lower panel)
as a function of blackbody luminosity (see text).}
\end{figure}

\begin{figure}[!ht]                                                           
\includegraphics[width=54mm,height=64mm,angle=270]{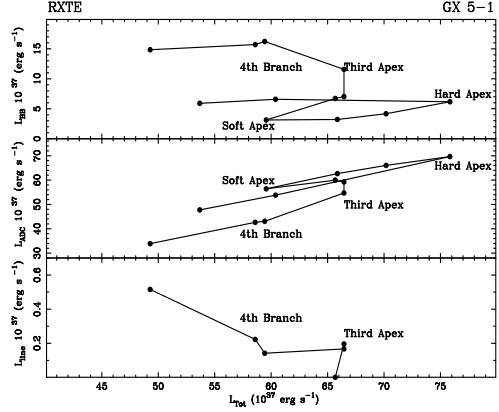}
\caption{Evolution of the neutron star blackbody luminosity (upper); the ADC Comptonized emission
(middle) ; the emission line (lower).}
\end{figure}

It can be seen in Fig. 5 that this unfolds the data and makes it clear that the behaviour
on the 4th branch is a continuation of that seen on the FB, i.e. 
the 4th track is a continuation of flaring, which becomes very strong on the 4th track, i.e.
there is a strong increase in the blackbody emission of the neutron star. The emission
has increasing area, and on the 4th track $R_{\rm BB}$ becomes very much greater than
the neutron star radius, rising to a maximum of 21 km. 

In Fig. 6 we show the variation of the luminosity components: the neutron star blackbody
luminosity $L_{\rm BB}$ (upper panel) and the luminosity of the ADC Comptonized emission 
$L_{\rm ADC}$ (middle panel) with $L_{\rm Tot}$. There is a strong increase 
of $L_{\rm BB}$ on the flaring branch which is continued on the 4th branch. However, at the 
third apex, $L_{\rm ADC}$ has begun to decrease and on the 4th branch decreases strongly.
The lower panel shows the strong emission feature seen on the 4th branch modelled
as RRC emission (below); this rose to 15\% of the neutron star luminosity or 4\% of 
the total luminosity.

The reason for the down-curving of the 4th track thus becomes
clear: the overall luminosity decreases 
because the Comptonized ADC emission is decreasing. Because $L_{\rm ADC}$ is $\sim$90\% 
of the total luminosity (at the soft apex), the decrease on the 4th track is a firm
indicator that the mass accretion rate $\dot M$ decreases. Thus the 4th track is simply 
the occurrence of strong flaring, i.e. unstable nuclear burning, under conditions of 
decreasing $\dot M$.

The spectra on the 4th branch clearly displayed strong broad emission which could be
well-fitted by addition of a Gaussian line. In the 4 spectra spaced along
the 4th track beginning at the third apex, the line energies were 9.49$\pm 0.67$, 
9.72$\pm 0.57$, 8.80$\pm 0.25$ and 7.58$\pm 0.78$ keV. The equivalent width 
increased from 320 eV to 1480 eV along 
the 4th branch. However, these energies are not consistent with Fe K emission but are 
suggestive of Fe radiative recombination continuum (RRC) characteristic of the various edges 
of highly ionized Fe. Consequently, we fitted the emission using the {\sc redge} model 
in {\it Xspec}. This model produced equally good quality fits; results are shown in 
Table 1 and a typical fit in Fig. 7.

\vskip -4mm
\tabcolsep 1.2 mm
\begin{table}[!h]
\begin{center}
\caption{Results of fitting the {\sc redge} model}
\begin{tabular}{rrr}
\hline \hline\\
spectrum & edge energy & $kT_{\rm e}$ \\
& (keV) & (keV)\\
\noalign{\smallskip\hrule\smallskip}
9  & 8.98$\pm$0.27 & 1.03$\pm$0.28 \\
10 & 9.39$\pm$0.22 & 0.76$\pm$0.22 \\
11 & 8.58$\pm$0.17 & 0.86$\pm$0.15 \\
12 & 7.77$\pm$0.09 & 1.13$\pm$0.13 \\
\noalign{\smallskip}\hline
\end{tabular}\\
\end{center}
\end{table}

\vskip - 8mm
In this model the free parameters are the edge energy, the line normalization and the temperature
$kT_{\rm e}$ of the recombining electrons which determines the width. The
edge energies associated with recombination of highly ionized states of Fe
to the next lower state are shown in Table 2 (Kallman 2003).

\vskip - 4mm
\begin{table}[!h]
\begin{center}
\caption{Edge energies for RRC in Fe}
\begin{tabular}{llll}
\hline \hline\\
edge energy & state   &edge energy & state\\
(keV) && (keV)\\
\noalign{\smallskip\hrule\smallskip}
9.278         &Fe XXVI     &8.088          &Fe XIX  \\   
8.755         &Fe XXV      &7.989          &Fe XVIII  \\
8.621         &Fe XXIV     & 7.891         &Fe XVII  \\
8.482         &Fe XXIII    &  7.838        &Fe XVI  \\
8.384         &Fe XXII     & 7.788         &Fe XV  \\
8.286         &Fe XXI      & 7.737         &Fe XIV \\
8.187         &Fe XX       & 7.686         &Fe XIII   \\
\noalign{\smallskip}\hline
\end{tabular}\\
\end{center}
\end{table}

\vskip - 8mm
The detected emission in spectrum 10 implies RRC emission 
of Fe XXVI recombination falling at the end
of the 4th branch to Fe XI. At any stage a mixture of states will probably
be present, and spectrum 12 at the flare peak was not well fitted with a single RRC,
providing some evidence of this.
In this case the fitting results may not give the electron temperature.

\begin{figure*}
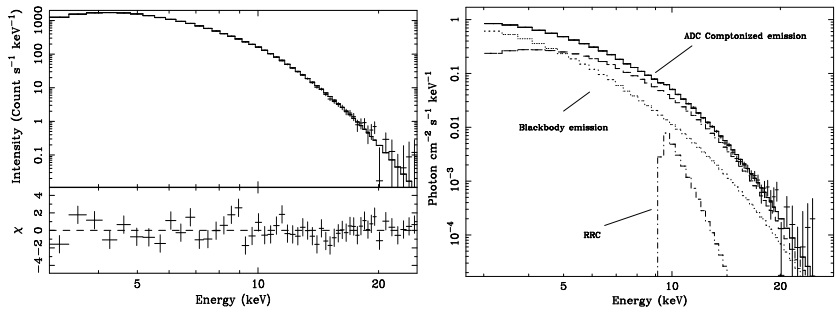

\begin{center}                                                            
\includegraphics[width=40mm,height=54mm,angle=270]{balucinska-church_2009_02_fig07.ps}                  
\includegraphics[width=40mm,height=54mm,angle=270]{balucinska-church_2009_02_fig08.ps}                 
\caption{Modelling of the emission feature at $\sim$9.5 keV in spectrum 10 with RRC emission: left: folded
spectrum with residuals; right: the unfolded spectrum.}
\end{center}
\end{figure*}

The presence of the emission only in flaring clearly shows it is associated with
flaring which we have suggested is unstable nuclear burning. It becomes strong
on the 4th branch where flaring is very strong. The increase in strength may well be due
to the increase of blackbody radius of the neutron star emission. Thus if there
is unstable nuclear burning that creates a fireball around
the neutron star extending to a radius of $\sim$21 km (Fig. 3) there will be a large volume in which 
the line can be generated. The decrease of edge energy towards the end
of the 4th branch correlates with the fall of $kT_{\rm BB}$ implying an eventual cooling.

\section{Discussion}

Our analysis of GX\th 5-1 shows that the 4th branch is simply a continuation of
the flaring branch in which the behaviour of the neutron star blackbody temperature
$kT_{\rm BB}$ and radius $R_{\rm BB}$ continues on the 4th branch as on the flaring branch, 
as shown in Fig. 4. On the 4th branch, $L_{\rm ADC}$ decreases showing that $\dot M$
decreases, so that the increase of $L_{\rm BB}$ can only be due to nuclear burning on 
the surface of the neutron star. We previously showed that in the three Cygnus\th X-2 
like sources (e.g. Church et al. 2006), at the soft apex, the value of $\dot m$, the 
mass accretion rate per unit area of the neutron star, agreed well with the critical value
below which nuclear burning is unstable (e.g. Bildsten 1998). Thus as the source moves
along the NB to the soft apex, unstable burning starts and the FB is formed.
The presence of a 4th branch shows that $\dot M$ continues to fall
and the 4th branch is a combination of unstable burning and this.

The observation of strong emission at energies between 7.8 and $\sim$9.4 keV is of great interest.
We suggest that this is RRC emission of ionization states Fe XXVI and lower
levels. Recombination of H-like Fe to He-like Fe produces RRC at 9.28 keV consistent with the
peak emission energy observed. The emission is very strong
which may be due to the large volume of
emitting plasma  around the neutron star suggested by the large values of blackbody radius
of up to 21 km. The decrease of line energy towards the end of the 4th track indicates
cooling as does the fall of blackbody temperature at the end of the 4th track.

The detection of Fe RRC is very rare in Galactic sources. However, a
broad line at $\sim$10 keV was previously seen in GX\th 5-1 on the flaring and 4th branch 
by Asai et al. (1994), probably the same feature as we have detected. They considered possible
origin of the line as Fe K$\alpha$, shifted to higher energies, either by a Doppler
shift in the disk, or by Compton scattering or by blue-shifting in a jet, but none of
these could provide a satisfactory explanation. 

A strong, broad feature at 9.28 keV has previously been seen in gamma ray bursts.
In GRB\th 970828, a broad emission feature at 4.8 keV
corrected for the well-determined redshift of the host galaxy of z = 0.9578 corresponds
closely to 9.28 keV (Yoshida et al. 2001). $kT_{\rm e}$ in this case was 0.8 keV
producing a very broad feature. Weth et al. (2000) considered conditions under
which the emission at 9.28 keV can be strong. In Galactic sources, narrow RRC 
emission of O VIII and O VII was seen in the dipping source XBT\th 0748-676
(Cottam et al. (2001) and in planetary nebulae (Nordon et al. 2009) 
in which $kT_{\rm e}$ is only a few eV.

In our case, the peak energy on the 4th branch corresponds to Fe XXVI RRC, i.e. H-like iron
and it appears that this emission is associated with the fireball 
around the neutron star resulting from uncontrolled nuclear burning. 
Further detailed study is clearly required.
 
\begin{acknowledgements}

This work was supported in part by the Polish grant 3946/B/H03/2008/34.
\end{acknowledgements}

\bibliographystyle{aa}

\end{document}